# IT STRATEGIC ALIGNMENT IN THE DECENTRALIZED FINANCE (DEFI): CBDC AND DIGITAL CURRENCIES.

**Carlos Alberto Durigan Junior** ; https://orcid.org/0000-0003-2185-493X
POLI USP

**Fernando José Barbin Laurindo** ; https://orcid.org/0000-0002-5924-3782
POLI USP

**IT Strategic alignment in the decentralized finance (DeFi): CBDC and digital currencies.**

## Abstract

Cryptocurrency can be understood as a digital asset transacted among participants in the crypto economy. Every cryptocurrency must have an associated Blockchain. Blockchain is a Distributed Ledger Technology (DLT) which supports cryptocurrencies, this may be considered as the most promising disruptive technology in the industry 4.0 context. Decentralized finance (DeFi) is a Blockchain-based financial infrastructure, the term generally refers to an open, permissionless, and highly interoperable protocol stack built on public smart contract platforms, such as the Ethereum Blockchain. It replicates existing financial services in a more open and transparent way. DeFi does not rely on intermediaries and centralized institutions. Instead, it is based on open protocols and decentralized applications (Dapps). Considering that there are many digital coins, stablecoins and central bank digital currencies (CBDCs), these currencies should interact among each other sometime. For this interaction the Information Technology elements play an important whole as enablers and IT strategic alignment. This paper considers the strategic alignment model proposed by Henderson and Venkatraman (1993) and Luftman (1996). This paper seeks to answer two main questions 1) What are the common IT elements in the DeFi? And 2) How the elements connect to the IT strategic alignment in DeFi? Through a Systematic Literature Review (SLR). Results point out that there are many IT elements already mentioned by literature, however there is a lack in the literature about the connection between IT elements and IT strategic alignment in a Decentralized Finance (DeFi) architectural network. After final considerations, limitations and future research agenda are presented.

## 1. Introduction and Objectives

Decentralized Finance (DeFi) stands for decentralized applications (Dapps) providing financial services on a blockchain settlement layer that is able to include payments, lending, trading, investments, insurance, and asset management (Defi Pulse, 2021) DeFi is able to cover a variety of activities relationships.  like stablecoins, exchanges, credit, derivatives, insurance among others. DeFi operates in a decentralized environment (public, permissionless blockchains). Services are generally encoded in open-source software protocols and smart contracts. DeFi protocols seek to disintermediate finance in a new governance. The market experienced explosive growth beginning in 2020. According to tracking service Defi Pulse DeFi grew over $15 billion at the end of 2020, and over $80 billion in May 2021(Defi Pulse, 2021).

DeFi is mainly based on the application of blockchain beyond cryptocurrency what generally involves private or permissioned blockchains that are controlled by a central entity or consortium of entities that governs the information flown among participants. According to Van der Merwe (2021), the crypto economy typically consists of four, interrelated components, as follows: I. The distributed ledger or blockchain, II. Digital assets, III. The active participants or miners and IV. The passive participants or users. Blockchain is a technology that allows a growing list of data structures (blocks) connected and secured by cryptography (Haber and Stornetta, 1990). A particular Blockchain is composed of blocks or groups of cryptocurrency transactions (Van der Merwe, 2021).

According to Zetzsche *et al* (2020) there is the "ABCD" of DeFi, this stands for the main four technologies which support DeFi, these are AI, Blockchain (including distributed ledgers and smart contracts), Cloud, and Data (big and small); or, in another iteration, AI, Big Data, Cloud, and DLT (including blockchain and smart contracts) (Zetzsche *et al*, 2020). One of the major applications of DeFi incentive structures is governance. Tokens issued in connection with liquidity mining or related mechanisms often provide governance rights for the DeFi service (Wharton , 2021).

|According to Hsieh et al (2017) cryptocurrencies also have their own governance aspects (both internal and external) While the effectiveness of internal governance is mainly based on the design of incentives, the effectiveness of external governance depends on the influence exerted by external factors and players. Since March 2017, the cryptocurrency market has become increasingly competitive, Bitcoin does not govern the market alone there has been a growth of a more diverse range of blockchain-based governance models, which entail additional complexity relative to traditional corporate governance. These new forms of governance, which are centralized in computer codes, emphasize the need for new research on organizational governance accounting for the interdependence of various levels about blockchain-based organizations. Foreseeably, this collaboration between centralized financial institutions and decentralized blockchain organizations will also foster the emergence of hybrid governance forms across organizational boundaries (Hsieh *et al*., 2017).

Convertibility among monetary instruments and interoperability between platforms will be crucial in reducing barriers to trade and enabling competition. Digital currencies may also cause an upheaval of the international monetary system: countries that are socially or digitally integrated with their neighbors may face digital dollarization, and the prevalence of systemically important platforms could lead to the emergence of digital currency areas. The advent of digital currencies will have implications for the treatment of private money, data ownership regulation, and central bank independence. For monetary policy to influence credit provision and risk sharing. In a digital economy where most activity happen through networks with their own monetary instruments, a regime in which all money is convertible to a central bank digital currency (CBDC) would uphold the unit of account status of public money, if a CBDC worked like stablecoins (Auer & Böhme, BIS 2021).

Laurindo (2008) stated that Information Technology (IT) is a widely accepted term that includes in its meaning; equipment (such as computers, servers, network, communication technology, automation, and network devices), applications, services, human, administrative and organizational aspects (Laurindo, 2008; Porter & Millar, 1985). IT plays an important function in leading with business processes and activities.

According to Luftman (2000) Strategic alignment is related to a managerial activity that should achieve cohesive goals across the Information Technology (IT) and other functional organizations. The IT and businesses functions should have their strategies well adapted together, alignment is evolutionary and dynamic demanding good managerial actions, communication and corporate commitment. There are two main directions in the IT strategic alignment: I. IT alignment with the business and II. Business aligned with IT (Luftman, 2000).

Luftman (1996) defined the twelve components of the strategic alignment model. these components set relationships that exist among them and IT, influencing the alignment of both directions. Furthermore (Luftman *et. al*, 1999) pointed out the enablers and inhibitors related to the strategic alignment. Henderson and Venkatraman (1993) state that IT, solely and exclusively, is not a source of competitive advantage, however it can be used in an aligned manner with organizational strategy (Henderson and Venkatraman, 1993).

Considering this introduction, it is justifiable to explore and comprehend further the roles that Information Technology (IT) plays in the universe of DeFi. In the crypto market there are many different digital currencies. There might exist the need of convertibility and comparability among these coins. Considering the features of stablecoins or a central bank digital currency (CBDC) there may be the need of comparing an equivalent value among these coins (backed or referred to stablecoins or CBDC). Furthermore, there should be some common IT features (or enablers) that allow convertibility among digital currencies. It is also important to consider that DeFi is set on a shared IT infrastructure and network, therefore different systems may connect or interact among themselves.

In this context this paper seeks to answer to main questions: *1) What are the common IT elements in the DeFi? And 2) How the elements connect to the IT strategic alignment in DeFi?* Figure 1 shows interactions among cryptocurrencies, as each one has its own Blockchain / DLT system. One possible interaction is interoperability among platforms, for this it is necessary to exist some IT resources and alignments. In the first question the term "*element*" can be understood as; tool, enablers, functions, resources, skills and features. This paper does not bring the literature definition of each element found.

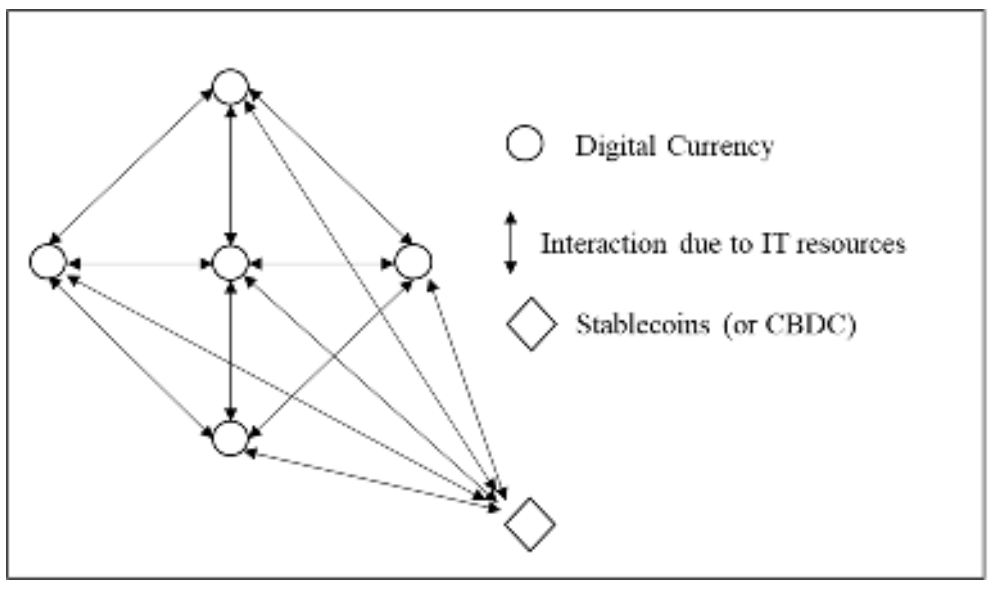

**Figure 1.** Digital currencies interactions.
Source: Author.

Methodologically, this research relies on a Systematic Literature Review (SLR) based on (Tranfield, Denyer, & Smart, 2003) in combination with Kitchenham (2004) and Kitchenham *et al*. (2009). The theoretical background adopted is Henderson and Venkatraman (1993) and Luftman (2000), both related to IT strategic alignment. Following this introduction, this paper is composed by a literature review, theoretical background, methodology, results and discussion, and finally conclusions, limitations and future agenda are presented.

## 2. Literature Review

### 2.1 Blockchain:

Blockchain is a technology that allows a growing list of data structures (blocks) connected and secured by cryptography (Haber and Stornetta, 1990). In the Blockchain, the distribution of information is decentralized, therefore Blockchain has been a technology able to provide decentralization, immutability, and transparency. Bitcoin, a digital currency is the first successful attempt to apply the technology (Satoshi Nakamoto, 2008).

The World Economic Forum (WEF) published a report in August 2016 named "The future of financial infrastructure an ambitious look at how Blockchain can reshape financial services". Blockchain systems use a Distributed Ledger Technology (DLT) and the WEF focused on topics in the financial markets where this technology can be applied to optimize the process and reduce costs. For instance, trade finance, global payments, and assets clearing/settlement are some evidences where this technology can be considered (WEF, 2016).

Blockchain-based solutions work well for both payments and settlement frameworks, using decentralized protocols. It is possible to make international payments and combine any currency. Transactions can be settled directly between the parties. According to Tasca (2016) Bitcoin seems to have been used more to transfer a large amount of money from person to person rather than used as payment for general consumptions (Tasca, 2016).

Considering governance, both legal code and technical code (software/hardware) may regulate general aspects. The impact of both must be considered in setting out regulations that cover distributed ledger systems. Lehdonvirta and Ali (2016) pointed out distinctions between governance (rulemaking by the owners or participants of a system to safeguard their private interests) and regulation (rulemaking by an outside authority tasked with representing the interests of the public) (Lehdonvirta and Ali, 2016).

Topics related to Information Technology, such as computing and cryptography have led to infrastructures that allow disintermediated and decentralized markets. Tasca (2015) declares that there are at least nine possible mechanisms into which Blockchain technology might be considered: 1) intermediation; 2) clearing and settlement; 3) recording systems; 4) rating or voting systems; 5) database systems; 6) distributed storage, authentication, anonymization of private information; 7) rewarding and punishing incentive schemes; 8) transaction traceability schemes; and 9) refereeing, arbitration, or notarization (Tasca, 2015).

### 2.2 Cryptocurrencies:

According to Tasca (2015), cryptocurrencies can be defined as: *"Money expressed as a string of bits sent as a message in a network that verifies the authenticity of the message via different mechanisms, such as proof-of-work (PoW) or proof-of-stake (PoS)"*. For accountability, every transaction needs to be transparent. Anonymity is preserved, however, all transactions are traceable regarding the fact that they are recorded in a public ledger (Tasca, 2015).

Tasca (2016) points out that economic theory defines money by looking at its functions as a medium of exchange, a unit of account, and a store of value. The author adds a fourth monetary aspect which was named as "*Transactional utility of reward*". This last aspect is related to the current digital era, which is characterized by a cashless and massively connected society utilizing high-frequency transnational transactions of products and services that have been being more digitalized. The transactional utility of reward gains relevance supported by global digital wallets (Tasca, 2016).

Cryptocurrencies are decentralized digital currencies; Bitcoin is the most widely known one. In more detail, a cryptocurrency is a digital token that exists within a particular system which generally consists of a P2P network, a consensus mechanism, and a public key. A cryptocurrency has three main properties: Digital barrier asset, Integrated payment network, and also non-monetary use cases.

There are many interdependencies in economic systems, they involve from simple local transactions to global large investment networks, single or clustered. The systemic complexities of economic networks depend not only on micro factors but also are influenced by macroeconomic forces. In the literature, there are two principal approaches to study economic networks: socioeconomics and complex systems (Schweitzer et. al, 2009). For instance, Bitcoin is a tiny fraction of the global economy, yet its network can disrupt the existent global economic governance model (Carlson, 2016).

According to the Bank for International Settlements (BIS), if Central Banks use cryptocurrencies to consumers and firms, this could significantly affect core banking areas such as payments, financial stability, and monetary policy (BIS, 2018).

### 2.3 Decentralized Finance (DeFi)

According to Schar (2021) Decentralized finance (DeFi) is a blockchain-based financial infrastructure. The term refers to an open, permissionless, and highly interoperable protocol on public smart contract platforms, such as the Ethereum blockchain. DeFi does not rely on intermediaries and centralized institutions. Instead, it is based on open protocols and decentralized applications (DApps). Agreements can be enforced by using code, transactions are executed under secure and verifiable procedures keeping legitimate state on public blockchains. Thus, this architecture is able to set an immutable and highly interoperable financial system with unprecedented transparency, equal access rights, existing minimum need for custodians, central clearing houses or other intermediaries. Smart contracts can perform roles executed by intermediaries. “smart contracts.” (Schar, F., 2021).

Smart contracts refer to applications stored on a blockchain and executed by a set of validators. For public blockchains, the network is designed allowing each participant to be involved and verify the correct execution of any operation. Smart contracts are somewhat inefficient compared with traditional centralized computing. However, their advantage is a high level of security. DeFi has the potential to set an open, transparent, and immutable financial infrastructure. Considering that DeFi consists of numerous highly interoperable applications (and protocols), everyone in the system can verify all transactions and data. DeFi leads to a more open and transparent financial infrastructure (Schar, F., 2021).

Due to blockchain technology and Ethereum (Distributed Ledger Technologies DLTs), market players became less dependent on intermediaries, moving the existing financial services to the blockchain and cryptocurrency environment. Ethereum is a blockchain-powered open software platform that gives developers opportunity to create and publish decentralized applications. The platform allows writing a code and using smart contracts that that runs a financial service and automates performances between two parties, eliminating the need of the third party. Smart contracts enables accuracy, transparency and security. Most of DeFi applications are developed in Ethereum environment using an open code. The users are able to combine many financial services, which open doors for making transactions in the network. (Stepanova and Riga 2021).

DeFi includes 3-D (I. Digitalization, II. Decentralization, and III. Democratization) concepts of sustainable development that motivate the users to employ the blockchain technology in financial services: I. Digitalization – refers to the use of digital technologies that influences economic spheres and communication on daily basis activities; II. Decentralization – there is the removal of the control played by intermediaries or large major players, there may be less transaction costs along with more interaction by participants in the network effect. ; III. Democratization – equal opportunities, financial services are accessible to anyone in the globe (Stepanova and Riga 2021).

DeFi comprises the following main elements: Digital assets and these include: a) Cryptoasset – intangible personal property used as a mean of exchange in the decentralized electronic environment and that does not have a tangible asset backing; b)Stablecoin – a cryptoasset characterized by stability of its value, which guards its owner from price volatility. The value of a stablecoin is referred to the value of another real asset; c) Global stablecoins – digital assets with characteristics of stable crypto coins that are created by large enterprises with a large user base. (Stepanova and Riga 2021).

### 2.4 CBDC and Stablecoins

Stablecoins can be understood as a category of cryptocurrency that seeks to stabilize the price by connecting the value to an underlying basket of assets. Stablecoins may work as digital equivalent of stable value funds, but their design is rather complex and involves the broader crypto-economy. Stablecoins may require a governing body, exchanges, wallet providers, payment system operators, smart contracts, and a DLT or blockchain system. Stablecoins are able to be backed by USD or other cryptocurrencies (crypto-collateral) (Van der Merwe , 2021).

According to the International Monetary Fund (IMF) Central Bank Digital Currency (CBDC) may be understood as a kind of legal tender in digital form, following the primary money functions (Kiff *et al*., 2020). CBDC is the evolution in money format from metal currency to metal-backed banknotes, and then to fiat money (Lee *et al*, 2021). An important difference between cash and electronic retail money is that the latter represents a claim on an intermediary, whereas the former is a direct claim on the central bank (BIS, 2020). A CBDC should allow central banks provide a universal means of payment for the digital era, safeguarding consumer privacy and preserving the private sector's primary role in retail payments and financial intermediation (BIS, 2020).

CBDC can foster competition among private sector intermediaries and serve as a basis for sound innovation in payments (Brunnermeier*, et al.* 2021). The advent of these new monies could reshape the nature of currency competition, the structure of the international monetary system, and the role of government-issued public money. In a digital economy, cash may effectively disappear, and payments may center around social and economic platforms rather than banks. Governments may need to offer central bank digital currency (CBDC) in order to retain monetary independence (Brunnermeier*, et al.* 2021).

## 3. Theoretical Background

Henderson and Venkatraman (1993) state that IT, solely and exclusively, is not a source of competitive advantage, however it can be used in an aligned manner with organizational strategy. The authors point out two basic pillars regarding strategic alignment, they are internal and external to the firms. For the authors, there is still a fundamental interaction between domains considering four essential factors; namely: 1-Business Strategy; 2-IT Strategy; 3-Infrastructure and Organizational Processes; and 4-IT Infrastructure and Processes.

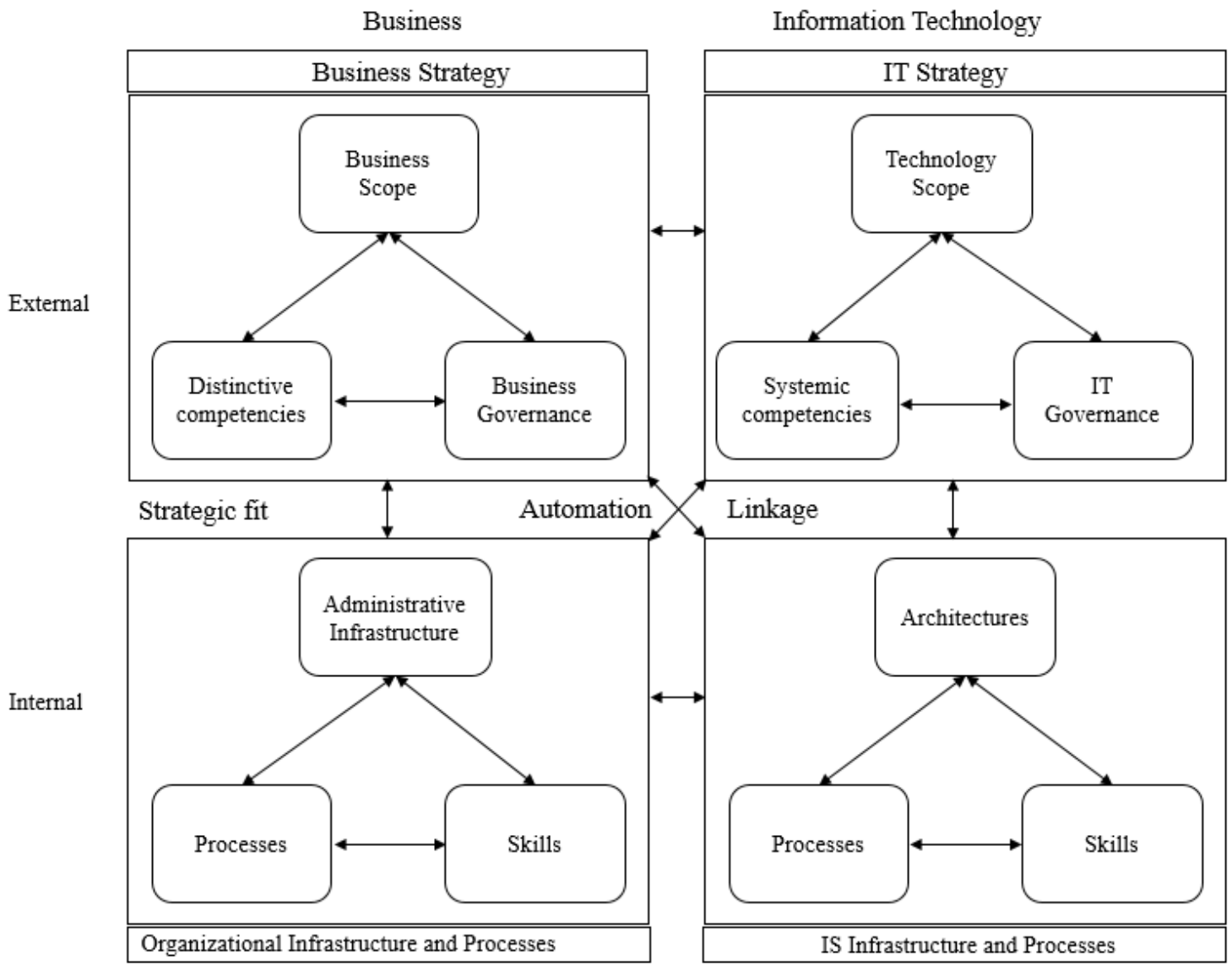

**Figure 2.** IT Strategic Alignment Model.
Source: Adapted from Henderson and Venkatraman (1993)

For Laurindo (2008), for each cycle of strategic alignment, three of the four domains are affected. According to the drivers, it is possible to have four alignment perspectives.

I. Strategy Execution: Refers to the impact that the business strategy generates directly on the business infrastructure, which in turn indirectly impacts the IT infrastructure; II. Technological Transformation: Refers to the impact that the business strategy generates directly on the IT strategy, which in turn impacts the IT infrastructure; III. Competitive Potential: Refers to the impact that the IT strategy generates directly on the business strategy, which in turn generates an impact on the organizational infrastructure; IV. Service level: refers to the impact that the IT strategy has directly on the IT infrastructure, which in turn has an impact on the organizational infrastructure. Figure 3 is from the alignment perspective.

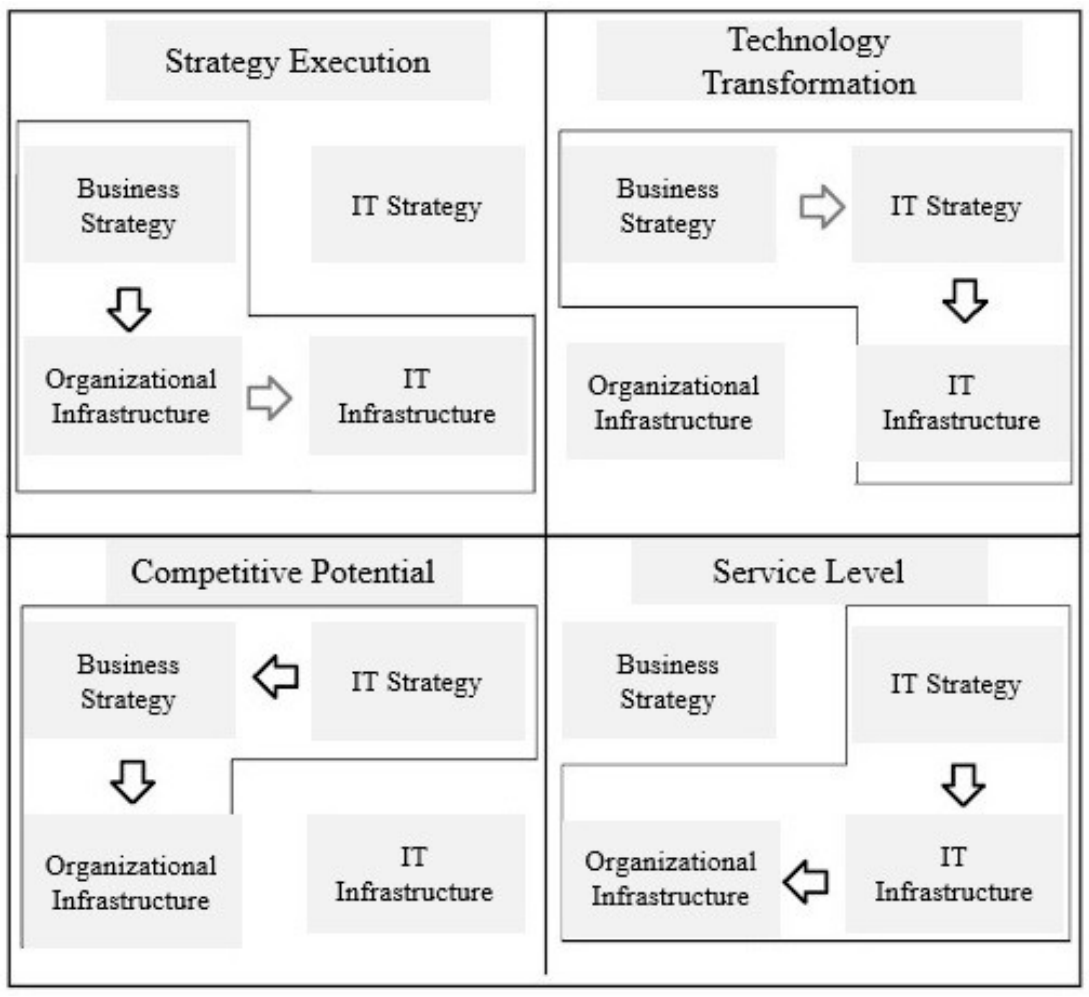

**Figure 3.** Strategic Alignment Perspectives of Henderson & Venkatraman (1993).
Source: Adapted from Laurindo (2008), Henderson & Venkatraman (1993)

Strategic alignment is related to a managerial activity that should achieve cohesive goals across the Information Technology (IT) and other functional organizations. The IT and businesses functions should have their strategies well adapted together, alignment is evolutionary and dynamic demanding good managerial actions, communication and corporate commitment. There are two main directions in the IT strategic alignment; I. IT alignment with the business and II. Business aligned with IT (Luftman, 2000).

IT Alignment's importance has been considered by scholars since the late 1970's (e.g., McLean & Soden, 1977; IBM, 1981; Mills, 1986; Parker & Benson, 1988; Brancheau & Whetherbe 1987; Dixon & Little, 1989; Niederman et al., 1991; Chan & Huff, 1993; Henderson, J., & Venkatraman, N. 1996; Luftman & Brier, 1999). According to the literature there are good evidences that IT has the power to transform whole industries and markets. (e.g., King, 1995; Luftman, 1996; Earl 1993; Earl, 1996; Luftman et. al., 1993; Goff, 1993; Liebs, 1992; Robson, 1994; Luftman, Papp, Brier, 1999; Luftman, Brier, 1999). Luftman (2000) presented an approach for assessing the maturity of a firm's business-IT alignment.

Luftman (1996) defined the twelve components of the strategic alignment model. these components set relationships that exist among them and IT, influencing the alignment of both directions. Table 1 shows.

Table 1
**The twelve components of strategic alignment**

| Category | Components |
|---|---|
| Business Strategy | 1. Business Scope |
| | 2. Distinctive Competencies |
| | 3. Business Governance |
| Organization Infrastructure and Processes | 4. Administrative Structure |
| | 5. Processes |
| | 6. Skills |
| IT Strategy | 7. Technology Scope |
| | 8. Systemic Competencies |
| | 9. IT Governance |
| IT infrastructure and Processes | 10. Architecture |
| | 11. Processes |
| | 12. Skills |

Source: Adapted from Luftman (1996).

Furthermore (Luftman *et. al*, 1999) pointed out the enablers and inhibitors related to the strategic alignment. Table 2 shows.

Table 2
**Enablers and Inhibitors related to the strategic alignment**

| | **Enablers** | **Inhibitors** |
|---|---|---|
| 1 | Senior executive support for IT | IT/business lack close relationships |
| 2 | IT involved in strategy development | IT does not prioritize well |
| 3 | IT understands the business | IT fails to meet commitments |
| 4 | Business-IT partnership | IT does not understand business |
| 5 | Well-prioritized IT projects | Senior executives do not support IT |
| 6 | IT demonstrates Leadership | IT management lacks leadership |

Source: Adapted from Luftman (1999).

Luftman (2000) points out six IT-business alignment maturity criteria, Table 3 shows.

Table 3
**IT-business alignment maturity criteria**

| | |
|---|---|
| Six IT Business Alignment Maturity Criteria | 1. Communications Maturity |
| | 2. Competency/Value Measurement Maturity |
| | 3. Governance Maturity |
| | 4. Partnership Maturity |
| | 5. Scope & Architecture Maturity |
| | 6. Skills Maturity |

Source: Adapted from Luftman (2000).

## 4. Methodology

Considering these questions this paper applies Sistematic Literature Review (SLR) approach based on Tranfield, Denyer, & Smart (2003) in combination with Kitchenham (2004) and Kitchenham *et al*. (2009). As suggested by these authors, the literature review can be subdivided into three main phases: planning the review, conducing the review, and reporting it.

Questions to be answered:

*1) what are the common IT elements in the DeFi? And;*
*2) How the elements connect to the IT strategic alignment in DeFi?*

I. Planning the Review

The review considers the following meanings: “Decentralized Finance” OR “DeFi” AND “Information Technology” AND “Strategic Alignment”. Sources considered: Scopus and Web of Science. These sources were considered because they are among the most used sources in academic environment. They do provide most of published literature. Table 4 details the criteria.

Table 4
**Including and Excluding Criteria**

| Including Criteria | Excluding Criteria |
|---|---|
| • Academic publications (mainly papers) which could bring some connection between Information technology and Decentralized Finance (DeFi). | • Papers that do not stablish relationship between DeFi and Information Technology or were not helpful to answer the two main questions from this work.<br>• Papers about other fields (oyher than digital economy) |

Source: Author.

II. Conducting the review

The search was performed using the Web of Science and Scopus scientific databases using the final strings in Table 5. Drawing on the methodological frameworks of Tranfield *et al*. (2003); Kitchenham (2004) and Kitchenham *et al*. (2009). For Scopus database the terms were searched in abstracts, titles, and keywords, without any other constraints. For Web of Science database, the strings were searched in "Topics". In this phase, the following articles information were exported: title, authors, abstract, publication year, keywords, source title, document type and language. Thus, papers exported metadata were saved on Microsoft Excel spreadsheets and the duplicated were eliminated. The available literature found was selected, inclusion and exclusion criteria were applied. The full articles selected were exported and the quality criteria were applied. Based on the full content of each selected article, the data extraction was subject of a critical analysis to seek literature answers for the two questions of this paper.

Table 5
**Database and Search Strings**

| Search ID | Scientific database | Search String |
|---|---|---|
| A | Web of Science | SEARCHED IN TOPICS<br><br>The review considers the following meanings: ("Decentralized Finance") OR ("DeFi") AND ("Information Technology") AND ("Strategic Alignment"). |
| B | Scopus | SEARCHED IN TITLE, ABSTRACT AND KEYWORDS<br><br>The review considers the following meanings: ("Decentralized Finance") OR ("DeFi") AND ("Information Technology") OR ("Strategic Alignment"). |
| C | Scopus | SEARCHED IN TITLE, ABSTRACT AND KEYWORDS |
| | | The review considers the following meanings: (“Decentralized Finance”). |

Source: Author.

Notes: In the search “B” it had been previously applied “and” in the last condition (and “strategic alignment”) however, it was found no publication. Considering this it was used “or” in the end, then twenty-two works were found, however none was considered. The search “C” was performed using only “Decentralized Finance”, then seventy works were returned.

III. Reporting the review

This item is better written in the next section of this paper, Results and Discussion.

The papers considered for analysis were completely read. After reading them, it was possible to extract and/or infer some possible elements related to Information Technology that connects with the two research questions proposed in this paper. In the first question the term “*element*” can be understood as; tool, enablers, functions, resources, skills and features. This paper does not bring the literature definition of each element found.

## 5. Results and Discussion

After applying the search criteria, fourteen papers were considered to address the two research questions. Eleven papers from Web of Science and three from Scopus. Table 6 points out.

Table 6
**Numbers found**

| Numbers Found | Web of Science | Scopus[1] |
|---|---|---|
| Total Itens found | 31 | 70 |
| Considered for reading[2] | 11 | 3 |

Source: Author.

**[1]Search C. [2]Reached the including criteria and excluding repeated.**

### 5.1 Question 1- what are the common IT elements in the DeFi?

The elements extracted from the papers read are shown in Table 7.

Table 7
**IT elements found in literature**

| Authors | Elements |
|---|---|
| Brühl (2021). | • Decentralized Applications (Dapps);<br>• Tokens;<br>• Smart Contracts Platforms;<br>• Web 3.0;<br>• Blockchain. |
| Zhao *et al* (2021). | • Ethereum network;<br>• Block synchronization protocol;<br>• Synchronization between nodes;<br>• Ethereum nodes;<br>• Transaction pools between nodes. |
| Zhou *et al* (2021). | • Blockchain consensus;<br>• Ethereum network. |
| Qiu *et al* (2019). | • Decentralized application games (Dapps games);<br>• Smart contracts;<br>• Programming languages for blockchain systems. |
| Kaal (2021). | • Incentive Design;<br>• Path Dependencies;<br>• On-Chain Governance;<br>• Legal Designs;<br>• Decentralized Autonomous Organizations (DAOs)<br>• Level of Decentralization. |
| Schär (2021). | • Architecture and the various DeFi building blocks;<br>• Token standards;<br>• Decentralized exchanges;<br>• Peer to Peer protocol;<br>• Asset tokenization;<br>• Layers (The settlement layer (Layer 1), the asset layer (Layer 2), the protocol layer (Layer 3), the application layer (Layer 4), the aggregation layer (Layer 5)). |
| Palina *et al* (2021). | • Liquidity-mining and governance mechanisms in DeFi protocols;<br>• Smart contracts;<br>• Tokens. |
| Perez *et al* (2021). | • Stablecoins;<br>• Governance Token Influence;<br>• Governance Token Risks; |
| | • Contagion Effects;<br>• Dynamics of incentive structures across different DeFi protocols. |
| Kutsyk *et al* (2020). | • Blockchain;<br>• Ethereum;<br>• Smart contracts;<br>• Internet of Things;<br>• Artificial intelligence;<br>• Convergence of Decentralized Autonomous Organizations (DAOs). |
| Stepanova, and Eriņš (2021). | • Cryptoasset ;<br>• Stablecoin;<br>• Global stablecoins. |
| Caldarelli and Ellul (2021). | • Oracle Problem in DeFi;<br>• Stablecoins. |
| Zetzsche *et al* (2020). | • AI, Big Data, Cloud, and DLT;<br>• Tech dependency;<br>• Tech risk;<br>• Data, reserve, and tech localization;<br>• RegTech and embedded regulation. |
| Ducrée *et al* (2021). | • Programmable Money;<br>• Oracles;<br>• Tokenization of Assets;<br>• Transaction Speed and Fees;<br>• Interoperability, Configurability, and Sustainability;<br>• Tokens;<br>• Publication, Peer Review, and Funding System. |
| Bartoletti *et al* (2021). | • Cryptographic protocol composition;<br>• Domain-specific languages. |

Source: Author.

According to the results in the table above, it is possible to observe that many authors bring common elements to the literature, from this, it is clear that these elements are related to the technological characteristics from distributed ledger technologies (as Blockchain) and or for some architectural or governance aspects in the digital finance. Skills and other features were found too.

It is already observable that the literature has well advanced in finding IT elements relates to the Decentralized Finance (DeFi) universe and that many of these elements are connected to the technological basis that do support DeFi infrastructure and operational aspects, such as crypto economic networks (Web 3.0), Distributed Leger Technologies (DLTs), smart contracts, tokens among others. Many of these elements are either connected or correlated in terms of the system functionality.

## 5.2 Question 2 – How the elements connect to the IT strategic alignment in DeFi?

This research question is an attempt to establish a connection between the IT elements found in question one (Q1) with the theoretical background stated by Luftman (1996) and Henderson & Venkatraman (1993). It is important to mention that some elements can be associated with more than one category or all of them. Table 8 shows the connection between IT elements and Luftman components of Strategic alignment model.

Table 8

**IT elements and the components of strategic alignment**

| Category | Components | Elements Found Q1 |
|---|---|---|
| Business Strategy | 1. Business Scope | • Cryptoasset ;<br>• Stablecoin;<br>• Global stablecoins.<br>• Oracle Problem in DeFi; |
| | 2. Distinctive Competencies | |
| | 3. Business Governance | |
| Organization Infrastructure and Processes | 4. Administrative Structure | • Blockchain;<br>• AI, Big Data, Cloud, and DLT;<br>• Ethereum;<br>• Smart contracts;<br>• Internet of Things;<br>• Artificial intelligence;<br>• Convergence of Decentralized Autonomous Organizations (DAOs).<br>• Interoperability, Configurability, and Sustainability; |
| | 5. Processes | |
| | 6. Skills | |
| IT Strategy | 7. Technology Scope | • Incentive Design;<br>• Path Dependencies;<br>• On-Chain Governance;<br>• Legal Designs;<br>• Decentralized Autonomous Organizations (DAOs)<br>• Level of Decentralization.<br>• Tech dependency;<br>• Tech risk;<br>• Data, reserve, and tech localization;<br>• RegTech and embedded regulation.<br>• Blockchain consensus;<br>• Ethereum network. |
| | 8. Systemic Competencies | |
| | 9. IT Governance | |
| IT infrastructure and Processes | 10. Architecture | • Decentralized Applications (Dapps);<br>• Smart Contracts Platforms;<br>• Layers (The settlement layer (Layer 1), the asset layer (Layer 2), the protocol layer (Layer |
| | 11. Processes | |
| | 12. Skills | |
| | | 3), the application layer (Layer 4), the aggregation layer (Layer 5)).<br><br>• Web 3.0;<br>• Blockchain.<br>• Governance Token Influence;<br>• Governance Token Risks;<br>• Contagion Effects;<br>• Dynamics of incentive structures across different DeFi protocols.<br>• Cryptographic protocol composition;<br>• Domain-specific languages<br>• Interoperability, Configurability, and Sustainability;<br>• Tokenization of Assets;<br>• Transaction Speed and Fees; |

Source: Author.

In practical terms, competitiveness in decentralized finance (DeFi) could be seen or measured as a network competitiveness, once there are many connected players in a DeFi system. The components of strategic alignment model are related to competitiveness. The correlation presented in the table above is an attempt to connect the elements with the category of alignment, however most of them can be related to all categories. For instance, Blockchain influences both governance and architectural aspects, the level of decentralization influences both governance and strategy. An interesting point is the consideration of the interaction among crypto currencies and central bank digital currencies (CBDCs) and stablecoins (see figure 1). There are many operational and architectural elements that must be aligned in order to set operablefunctions among different DeFi platforms and protocols.

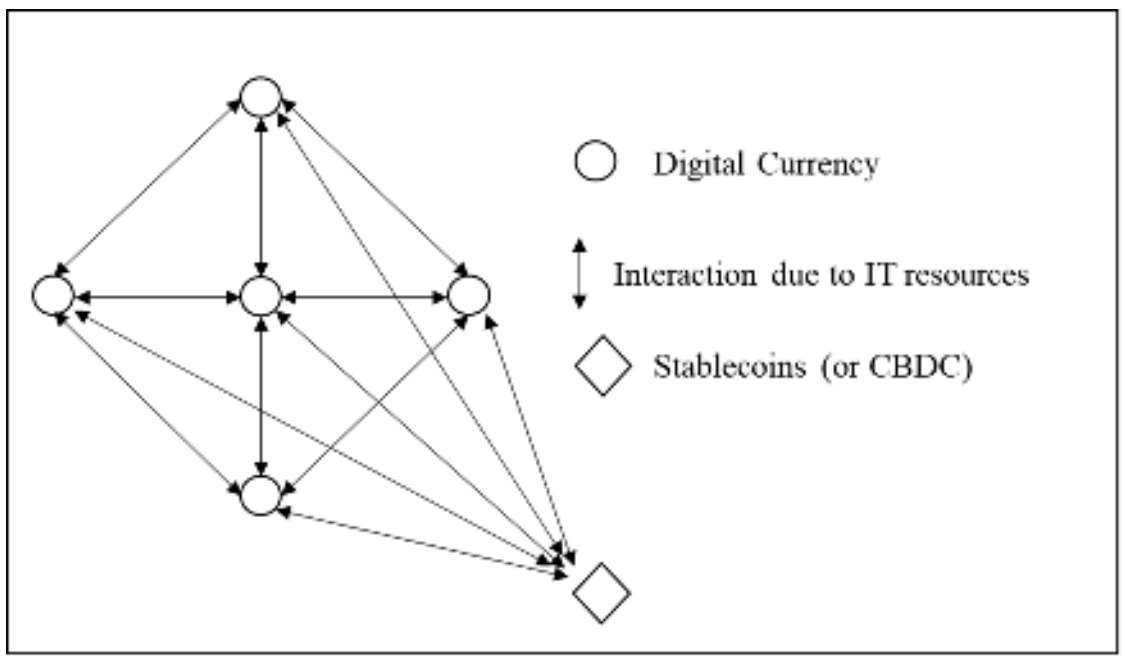

**Figure 1.** Digital currencies interactions.
Source: Author.

Considering the four alignments categories by Luftman and the IT elements associated in Table 8 we can comprehend:

- **Business strategy:** The functions of the coin in the DeFi system may influence the Business Strategy. If the coin is an stablecoin, a global stablecoin or has any other feature. Additionally, operational problems may limit the strategy.
- **Organization Infrastructure and Processes**: Skills, resources, employed technologies and features may influence all the governance of processes and the organizations in the DeFi system.
- **IT Strategy:** In this case, the IT Strategy may influence the strategy of all DeFi network, mainly when there are some factors related to tech dependencies, controlled level of decentralization, technological risks, regulation and network interactions.
- **IT infrastructure and Processes:** The resources do influence strategy in this point, like the use of smart contracts, decentralized applications (Dapps), tokens, Blockchain layers, crypto economic platforms (Web 3.0) among others.

According to Henderson & Venkatraman Strategic Alignment Perspective (1993) – Figure 3, it is coherent to observe that the financial industry, specifically in the cases of Decentralized Finance (DeFi) and digital economy is in a “Technological Transformation” stage.

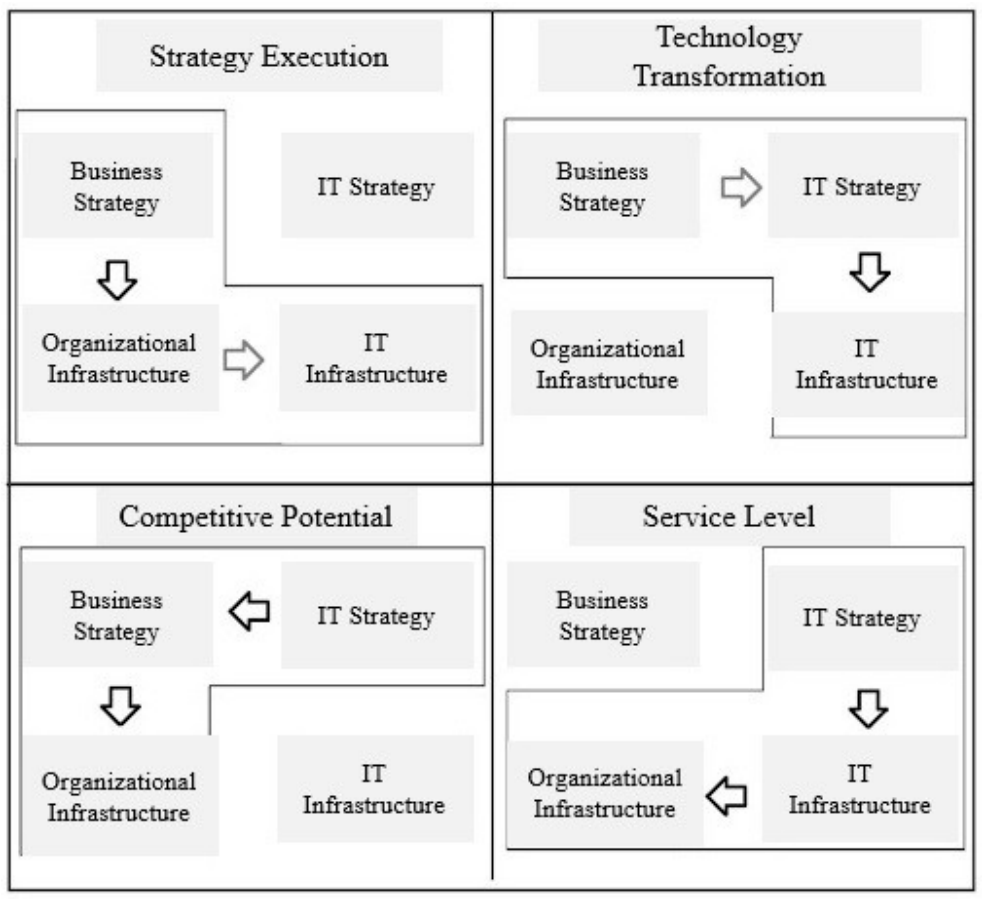

**Figure 3.** Strategic Alignment Perspectives of Henderson & Venkatraman (1993).
Source: Adapted from Laurindo (2008), Henderson & Venkatraman (1993)

## 6. Conclusions, Limitations and Future Agenda

This paper seeks to answer two main questions about the Information Technology (IT) elements in the Decentralized Finance (DeFi) technological architecture. Specifically, the questions were: 1) What are the common IT elements in the DeFi? And 2) How the elements connect to the IT strategic alignment in DeFi? For the first question it was found that the literature has already mentioned many IT elements about crypto economics and Decentralized Finance aspects as mentioned in Table 7. However, there is a lack in the literature about directly mentioning the IT elements in DeFi in connection with IT strategic alignment approach.

This paper tried to associate the IT elements found (Table 7) with the components of strategic alignment by Luftman (Table 8), then it is possible to see that for some elements there is a connection among many categories of alignments (not only one category). Considering that there will be interaction among central bank digital currency (CBDC), stablecoins and crypto currencies in DeFi protocols, interoperability and other infrastructure or operational element may be used and seen as source of competitiveness and IT strategic alignment.

The fact that this research context is very new, could be not completely accepted for being one limiting factor in this attempt, however there is a lack in the literature associating IT elements with IT strategic alignment in DeFi protocols, this last observation is a limiting factor. In this research it was found no paper directly covering this approach. This paper contributes to the literature by being pioneer in establishing a first attempt of a connection between IT elements and IT strategic alignment in DeFi context.

Future research agenda should keep exploring more about the topics and the research questions considered in this paper. It would be interesting if more researchers could go deeper in the questions of a better comprehension of the IT elements both in the IT governance and in the IT strategic alignment of a Decentralized Finance (DeFi) protocol. One future research question could be which IT elements are more connected to IT strategic alignment? and second; what elements can be source of competitiveness? (Considering Henderson & Venkatraman (1993) and /or Luftman (1996, 1999, 2000). Once DeFi networks are complex and cover many agents, there might be the need of also exploring what really are the sources of competitiveness in IT Strategic alignment, even a new definition of this term may be considered by scholars in the context of Distributed Ledger technologies.